\documentclass[aps,prl,superscriptaddress,floatfix,twocolumn,showpacs]{revtex4-1}

\usepackage{amsmath,amssymb}
\usepackage{graphicx}
\usepackage{color}

\usepackage{dsfont}
\usepackage{ifthen}
\usepackage{extarrows}
\usepackage{feynmp}
\usepackage[pdftex, pdfstartview = {FitH}]{hyperref}
\usepackage{ifpdf}
\ifpdf
\fi

\newcommand{\nn}[0]{ \nonumber \\}
\newcommand{\lP}[0]{ \left( }
\newcommand{\rP}[0]{ \right) }
\newcommand{\lB}[0]{ \left[ }
\newcommand{\rB}[0]{ \right] }
\newcommand{\lC}[0]{ \left\{ }
\newcommand{\rC}[0]{ \right\} }

\newcommand{\ks}[0]{\mathrm{s}}
\newcommand{\x}[0]{\mathrm{x}}
\newcommand{\xc}[0]{\mathrm{xc}}
\newcommand{\kf}[0]{k_\mathrm{F}}
\newcommand{\rs}[0]{{r_\mathrm{s}}}
\newcommand{\kp}[0]{\kappa_\mathrm{+}}
\newcommand{\km}[0]{\kappa_\mathrm{-}}

\begin{document}

\title{A shortcut to gradient-corrected  magnon dispersion: exchange-only case}

\author{F. G. Eich}
\email[]{florian.eich@mpsd.mpg.de}
\affiliation{Max Planck Institute for the Structure and Dynamics of Matter, Luruper Chaussee 149, D-22761 Hamburg, Germany}

\author{S. Pittalis}
\affiliation{Istituto Nanoscienze, Consiglio Nazionale delle Ricerche, Via Campi 213A, I-41125 Modena, Italy}

\author{G. Vignale}
\affiliation{Department of Physics and Astronomy, University of Missouri, Columbia, Missouri 65211, USA}

\begin{abstract}
  Ab initio calculations of the magnon dispersion in ferromagnetic materials typically rely on the adiabatic local density approximation (ALDA) in which the effective exchange-correlation field is everywhere parallel to the magnetization.  These calculations, however, tend to overestimate the ``magnon stiffness", defined as the curvature of the magnon frequency vs. wave vector relation evaluated at zero wave vector.  Here we suggest a simple procedure to improve the magnon dispersion by taking into account gradient corrections to the ALDA at the exchange-only level. We find that this gradient correction always reduces the magnon stiffness. The surprisingly large size of these corrections ($\sim 30\%$) greatly improves the agreement between the calculated and the observed magnon stiffness for cobalt and nickel, which are known to be overestimated within the ALDA.
\end{abstract}

\maketitle

\section{Introduction} \label{SEC:intro}

Hardy Gross, in whose honor this article is written, led many important developments in the Density-Functional Theory (DFT) of magnetic materials. One of his important contributions, and an area of strong overlap between his interests and ours~\cite{EichGross:13}, was the recognition that non-collinear spin systems require exchange-correlation magnetic fields that are not necessarily parallel to the magnetization~\cite{CapelleGyoerffy:01}. These transverse exchange-correlation (xc) fields appear as soon as one goes beyond a naive local spin density approximation, and are clearly seen, for example, in the exact-exchange formulation, which Hardy spearheaded several years ago~\cite{SharmaGross:07}.

\begin{figure}[t]
\begin{center}
\includegraphics[width=.8\linewidth]{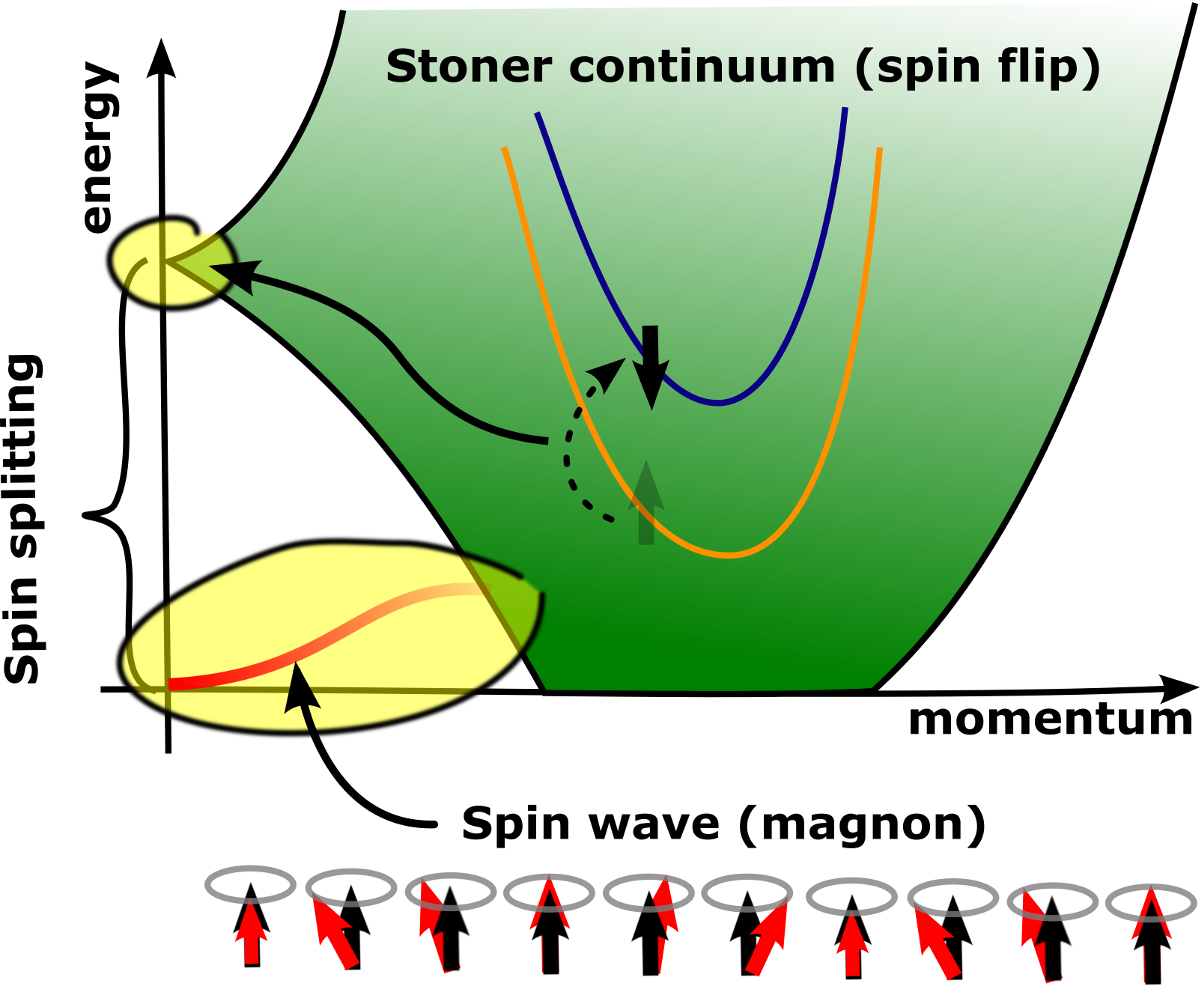}
\caption{Sketch of a typical transverse magnetic absorption spectrum based on a spin-polarized uniform electron gas (density of states of magnetic excitations). The Stoner continuum is shown as the green-shaded area and the dispersion relation of the magnons is highlighted as the red curve outside the Stoner continuum. For a ferromagnetic material--in the absence of spin-orbit coupling--the dispersion relation starts from zero energy at zero momentum, which reflects the fact that a global rotation of the spins does not change the energy of the ferromagnet (Goldstone mode).
}
\label{FIG:ferromagnet}
\end{center}
\end{figure}

This paper is devoted to the effect of transverse xc fields on the magnon stiffness of ferromagnetic materials. A magnon is an elementary excitation of a ferromagnetic material, characterized by a frequency $\omega_m$  and a wave vector $q$, which correspond to a collective precessional motion of the magnetization around its equilibrium value. In Fig.\ \ref{FIG:ferromagnet} we sketch a typical absorption spectrum of a ferromagnet. The green-shaded area corresponds to single spin-flip excitations, i.e., the so-called Stoner continuum. The red curve depicts the dispersion relation of the magnons. For small $q$ (small on the scale of the inverse of the lattice constant) the dispersion has the form
\begin{align}
  \omega_m(\vec{q}) \approx D q^2 ~, \label{omag}
\end{align}
which defines the ``magnon stiffness" $D$. We refer to the $q^2$-coefficient of the magnon dispersion as ``magnon stiffness" in order to make the distinction to the ``spin stiffness'' which is the \emph{minus} the inverse of the homogeneous spin susceptibility--a $q=0$ property. In contrast to the spin stiffness, which is always a positive quantity guaranteeing the stability of the system, the magnon stiffness may become negative, e.g., when the system is spin-polarized due to an external magnetic field.

A fully ab initio calculation of the magnon dispersion requires finding the poles of the $q$- and $\omega$-dependent spin-susceptibility, or, equivalently the zeros of its inverse $\Pi^{-1}(q,\omega)$.~\cite{Savrasov:98,KarlssonAryasetiawan:00,KotaniSchilfgaarde:08,BuczekSandratskii:09,SasiogluBluegel:10,LounisMills:11,RousseauBergara:12,CaoGiustino:18} The latter is conveniently represented  as the sum of an inverse Kohn-Sham susceptibility, which can be calculated by well established methods within static density functional theory, and an exchange-correlation kernel, which needs to be approximated:
\begin{align}
[\Pi^{-1}]_{\alpha\beta}(\vec{q},\omega)=[\Pi_\ks^{-1}]_{\alpha\beta}(\vec{q},\omega)-f_{\xc,\alpha\beta}(\vec{q},\omega)~,
\end{align} 
where the Greek indices refer to Cartesian coordinates of spin space. The so-called ALDA is the approximation in which the $q$- and $\omega$-dependence of $f_{\xc}$ is completely neglected, i.e., one sets $f_{\xc}(q,\omega) = f_{\xc}(0,0)$.  It is evident that this approximation neglects retardation effects in time (hence it is an adiabatic approximation), as well as non-locality in space (hence it is a local density approximation).  If spin-orbit interactions are neglected and external magnetic fields are absent, then the condition that the magnon frequency tend to zero for $q \to 0$ (Goldstone's theorem) completely determines the value of $f(0,0)$~\cite{MuellerBluegel:16}. This is the same as requiring that $[\Pi^{-1}]_{\alpha \beta}(0,0)$ vanishes (or equals the external magnetic field if one is present), i.e., the transverse spin susceptibility diverges at $q = 0$, implying a zero-frequency spin wave in the absence of any symmetry-breaking fields. 

The magnon dispersion calculated in this manner is found to produce a magnon stiffness that, at least in the case of the metallic ferromagnets cobalt and nickel, is substantially larger than the measured values. A possible reason for this overestimation is the neglect of the $q$-dependence of $f_{xc}$.  Continuing to work within the framework of the adiabatic approximation we can write
\begin{align}
  f_{\xc,\alpha\beta}(\vec{q},\omega) \simeq f_{\xc,\alpha\beta}(0,0)
  + \big[A_{\xc}\big]_{\alpha i, \beta j} q_i q_j ~, \label{fxcQexpand}
\end{align} 
which is valid when $q$ is much smaller than microscopic length scales such as the electronic Fermi wave vector and the inverse of the lattice constant (Latin indices correspond to Cartesian coordinates of the momenta, summation over repeated indices is implied). The $q$-dependent term introduces gradient corrections to the xc field (i.e., it goes beyond the local density approximation) and can thus modify the magnon dispersion at small $q$. In Eq.\ \eqref{fxcQexpand} (and throughout the paper) we assume inversion symmetry, which implies that there is no $q^1$ contribution to the response. The inverse of the \emph{ static} spin susceptibility can now be written as
\begin{align}
  \big[\Pi^{-1} \big]_{\alpha\beta} (\vec{q},0) \simeq \big[ \Pi^{-1} \big]_{\alpha\beta}(0,0) 
  - \big[A_{\ks} + A_{\xc} \big]_{\alpha i, \beta j} q_i q_j ~. \label{PiInvQ2}
\end{align} 
where $A_{\ks,\alpha i\beta j}$ is the coefficient of the small-$q$ expansion of the inverse Kohn-Sham response function
\begin{align}
  \big[ \Pi_\ks^{-1} \big]_{\alpha \beta}(\vec{q},0) \simeq \big[ \Pi_\ks^{-1} \big]_{\alpha\beta}(0,0)
  - \big[ A_{\ks} \big]_{\alpha i, \beta j} q_i q_j ~.
\end{align} 
The tensor, $A_{\alpha i, \beta j} \equiv \big[A_{\ks} + A_\xc \big]_{\alpha i, \beta j}$, determines the energy cost of a spatial variation of the spin according to the formula
\begin{align}
  \Delta E = \frac{1}{2} \big(\nabla_i \tilde{M}_{\alpha} \big) A_{\alpha i,\beta j}
   \big(\nabla_j \tilde{M}_{\beta}\big) ~, \label{DE}
\end{align}
where $\tilde{M}_\alpha \equiv M_\alpha - M^{(0)}_\alpha$ is the deviation of the magnetization from its equilibrium value $M^{(0)}_\alpha$. Formally, $A_{i \alpha,j \beta}$ can be identified as the (negative) coefficient of the quadratic term in the expansion of the inverse spin-spin response function in powers of the wave vector q as can be seen from Eq.\ \eqref{PiInvQ2}. This establishes a connection with the so-called  ``frozen-magnon'' approach \cite{HalilovOppeneer:98,GrotheerFaehnle:01,EssenbergerGross:11} to the calculation of magnon frequencies in which the magnetic moments of the ferromagnet are constrained to form a spin spiral (as shown at the bottom of Fig.\ \ref{FIG:ferromagnet}) and the energy difference to the ground state is calculated.

In many cases the magnitude of the magnetization can be assumed to be essentially constant, implying that a small change in magnetization $\tilde{M}_\alpha$ is everywhere orthogonal, i.e., transverse to, the equilibrium magnetization. Further, one can often reasonably neglect the relatively small anisotropy due to the spin-orbit coupling experienced by electrons in the lattice (crystalline anisotropy). With these simplifications, the tensor reduces to a single number $A$, such that
\begin{align}
  \Delta E = \frac{1}{2} A \big(\nabla_i \tilde{M}_{\alpha} \big) \big(\nabla_i \tilde{M}_{\alpha}\big) ~. \label{DE_A}
\end{align}
This number controls important material properties such as the width of domain walls and the magnon dispersion.  In particular, we will now show that the exchange correction to $A$ has a surprisingly large impact on the magnon stiffness.

The remainder of this paper is organized as follows: in Section~\ref{SEC:EGmagnons}, we derive the dispersion of magnons in a weakly inhomogeneous electron gas at the level of an exchange-only approximation; in Section~\ref{SEC:FMmagnons}, we exhibit the remarkable effect of gradient corrections on the magnon dispersion using a simple estimate based on the results for the uniform electron gas; in Section~\ref{SEC:conclusions}, we conclude and point out a possible way to capture more rigorously the effects of realistic magnetic inhomogeneities.

\section{Weakly inhomogeneous transverse spin fluctuations in the uniform electron gas} \label{SEC:EGmagnons}

Let us start by considering a uniform gas which is polarized along $z$--possibly by means of a magnetic field along $z$. The Zeeman term, coupling the spin magnetization to the magnetic field, contributes to the Hamiltonian with a term
\begin{align}
  \hat{H}_{\vec{B}} & = - \vec{\sigma} \cdot \vec{B} ~, \label{BsignDef}
\end{align}
and the expectation value of the vector of Pauli matrices, $\vec{\sigma}$, defines the spin magnetization, 
$\vec{m} = \langle \vec{\sigma} \rangle$. Within Spin Density-Functional Theory (SDFT) the magnetization of the interacting system is reproduces by a fictitious non-interacting (Kohn-Sham) system exposed to an effective (Kohn-Sham) magnetic field $\vec{B}_\mathrm{s} = \vec{B}_\mathrm{ext} + \vec{B}_\xc$. $\vec{B}_\mathrm{ext}$ corresponds to the magnetic field applied to the interacting system and $\vec{B}_\xc$ is the so-called xc magnetic field, representing an effective internal magnetic field due to the electron--electron interaction. The sign convention for the Zeeman term \eqref{BsignDef} and the decomposition of the Kohn-Sham magnetic field implies that the xc magnetic field is given by $\vec{B}_\xc = - \delta E_\xc / \delta \vec{m}$. In order to describe transverse magnetic fluctuations it is customary to introduce transverse magnetic fields $B_{\pm} = B_x \pm i B_y$~\cite{GiulianiVignaleLR:05}. These  fields couple to the transverse Pauli matrices $\sigma_{\pm} = \sigma_{x} \pm i \sigma_{y}$, respectively. The \emph{transverse} spin-spin response function, $\Pi_{+-} = 2 \big[ \Pi_{x,x} - i \Pi_{xy} \big]$~\footnote{Due to rotational symmetry we have $\Pi_{x,x}=\Pi_{y,y}$, furthermore we used $\Pi_{xy} = -\Pi_{yx}$.} contains the information of the transverse magnetic excitations in the system. The crucial point is that $\Pi_{+-}$ supports collective modes (magnons), which are transverse spin-fluctuations and are characterized by a dispersion relation $\omega_m(\vec{q})$. The dispersion relation is determined by the zeros of the inverse of $\Pi_{+-}(\vec{q},\omega)$.

Within time-dependent SDFT (TD-SDFT) the inverse of $\Pi_{+-}(\vec{q},\omega)$ can be expressed as follows:
\begin{align}
  \Pi^{-1}_{+-}(\vec{q},\omega) = \Pi_{+-, \ks}^{-1}(\vec{q},\omega) - f_{+-, \xc}(\vec{q},\omega) ~. \label{Pi_Pis_fxc}
\end{align}
In our work on the gradient expansion of the exchange energy, we have computed the long wavelength limit of the \emph{ static} exchange kernel $I_\x \equiv f_\x(\vec{q}, \omega=0)$~\cite{EichVignale:13}. More precisely, we have computed $I_{\perp, \x} \equiv I_{xx, \x} = I_{yy, \x}$, which differs from $I_{+-,\x}$ by a factor of $2$, i.e., $I_{+-,\x} = \frac{1}{2} I_{\perp, \x}$. Note that the $xy$-response vanishes in the adiabatic limit $\omega=0$, i.e., $\Pi_{xy}(\omega=0) = 0$. In the following, we drop the subscript ``$+-$'' for brevity. The long wavelength expansion for the static exchange kernel is given by (atomic units are used throughout the paper):
\begin{subequations}
  \begin{align}
    I_{\x}(\vec{q}) & \approx I_\x^{(0)} + I_\x^{(2)} q^2 ~, \label{IxExpansion} \\
    I_\x^{(0)} & = -\frac{\Delta_\x}{4 m} = \frac{\partial_m \epsilon_\x}{2 m}
    = - \frac{1}{4 m}\frac{\kf \lB \kp - \km \rB}{\pi}~, \label{Ix0} \\
    I_\x^{(2)} & =  \frac{D_\x}{4m} =   \frac{1}{4m} \frac{\frac{8}{9}p^2
    - \frac{2}{5}\lP \kp^5 - \km^5 \rP \lP \kp - \km \rP}{p \pi \kf \lB \kp^2 - \km^2\rB^2}
   ~. \label{Ix2} 
  \end{align}
\end{subequations}
The previous equations are written exclusively in terms of the density, $n$, and the spin magnetization, $m$, not to be confused with the electron mass which is one in atomic units. Moreover, we have introduced the Fermi wave vector $\kf = \lP 3 \pi^2 n \rP^{1/3}$, the relative spin polarization $p=m/n$ and the relative spin-up and spin down wave vectors $\kappa_{\pm} = \lP 1 \pm p\rP^{1/3}$. This means that the spin-up and spin-down Fermi wave vectors, $k_{\mathrm{F} \uparrow / \downarrow}$, are given by $k_{\mathrm{F} \uparrow } = \kf \kappa_+$ and $k_{\mathrm{F} \downarrow } = \kf \kappa_-$ and fulfill $\kf^3 = (k_{\mathrm{F} \uparrow}^3 + k_{\mathrm{F} \downarrow}^3) / 2$. $\Delta_\x = 2 B_\x = -2 \partial_m \epsilon_\x$ is the exchange contribution to the spin splitting energy determined from the derivative of the exchange energy density, $\epsilon_\x$, with respect to the magnetization. Note that $q$ is the wave vector in \emph{ atomic units} and not in units of $\kf$. Equation \eqref{Ix0} follows from the fact that the transverse (static) xc kernel of the electron gas is given by $I_{\perp, \xc}(\vec{q}=0) = m^{-1} \partial_m \epsilon_\xc$. The $q^2$-coefficient \eqref{Ix2} corresponds to half of Eq.\ (44) in Ref.\ \cite{EichVignale:13}, which provides details of its derivation.

The frequency-dependent Kohn-Sham response function for the \emph{ uniform electron gas} is given by~\cite{GiulianiVignaleLRlindhard:05}
\begin{align}
  \Pi_\ks(\vec{q}, \omega) & \approx \frac{4 m}{\omega - \Delta_\ks} + \frac{2 n}{\lP \omega - \Delta_\ks \rP^2} \lB 1 
  + \frac{ \kf^2 \lP \kp^5 - \km^5 \rP}{5 \lP \omega - \Delta_\ks \rP}\rB q^2 ~,
  \label{PisQ2}
\end{align}
again, in the long wavelength limit.\footnote{Note that Eqs.\ (4.14) and (E4.1) of Ref.\ \cite{GiulianiVignaleLRlindhard:05} are missing a factor of $2$, which has been corrected in the errata available at \url{http://faculty.missouri.edu/~vignaleg/books/}.} $\Delta_\ks = 2 B_\ks$ is the spin splitting energy of the Kohn-Sham system. Consequently, the magnon dispersion at small $q$ is determined by 
\begin{align}
  0 & = \Pi^{-1} = \Pi_\ks^{-1} - I_\x \label{OM} \\
  & \approx \frac{1}{4 m} \Bigg[ \omega - \Delta_\ks + \Delta_\x \nn
  & \phantom{\approx \frac{1}{4 m} \Bigg[ } {} - \lC \frac{1}{2 p} \lB 1
  + \frac{ \kf^2 \lP \kp^5 - \km^5 \rP}{5 \lP \omega - \Delta_\ks \rP} \rB + D_\x \rC q^2 \Bigg]
  ~. \nonumber
\end{align}
We can solve Eq.\ \eqref{OM} at $q=0$ which yields $\omega_m(q=0) = \Delta_\ks - \Delta_\x$. The solution at $q=0$ can be used to eliminate the $\omega$-dependence in the $q^2$ coefficient leading to the magnon dispersion
\begin{align}
  \omega_m(\vec{q}) \approx \Delta_\ks - \Delta_\x
  + \big( D_\ks + D_\x \big) q^2 ~. \label{OMq2}
\end{align}
In Eq.\ \eqref{OMq2} we have introduced
\begin{align}
  D_\ks & = \frac{1}{2p} \lB 1 - \frac{ \kf^2\lP \kp^5 - \km^5 \rP}{5 \Delta_\x} \rB \nn
  & = \frac{1}{2p} \lB 1 - \frac{\pi \kf \lP \kp^5 - \km^5 \rP}{5 \lP \kp - \km \rP} \rB
  ~, \label{Ds}
\end{align}
where in the second line we used the explicit expression for the exchange splitting $\Delta_\x$ given in Eq.\ \eqref{Ix0}. $D_\x$ has been defined in Eq.\ \eqref{Ix2}. Note that the Kohn-Sham contribution to the magnon stiffness, $D_\ks$, contains exchange effects ``in disguise'', since it depends on the exchange contribution to the spin splitting energy.

\begin{figure}[t]
  \begin{center}
    \includegraphics[width=0.9 \linewidth,angle=0]{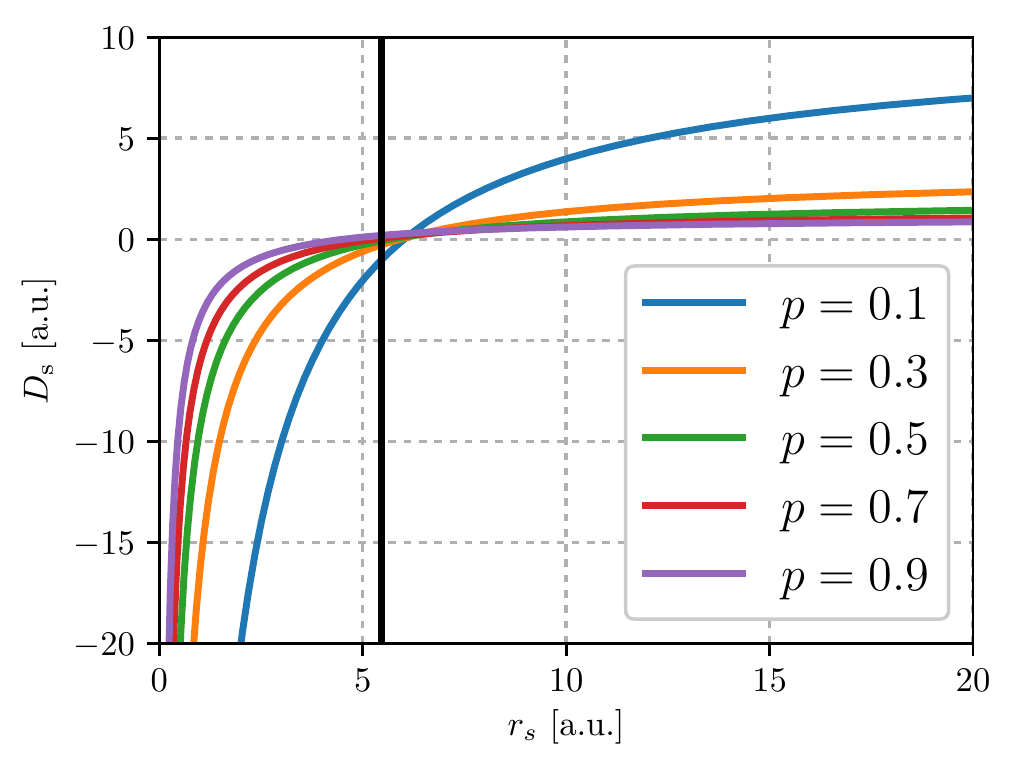}
    \caption{Plot of $D_\ks$ as function of $\rs$ for various relative spin polarization $p$. Note the change of sign as $r_s$ increases. The vertical black line corresponds to the critical $\rs$ above which the uniform electron gas is spontaneously--without an external magnetic field--fully spin polarized ($p=1$).}
    \label{FIG:Ds}
  \end{center}
\end{figure}
In Fig.\ \ref{FIG:Ds} we show $D_\ks$ as function of the density--given in terms of the Wigner-Seitz radius $\rs= \lP 9 \pi / 4 \rP^{1/3}\kf^{-1}$--for various relative spin polarizations $p$. At high densities (small $\rs$) $D_\ks$ is negative and at low densities (large $\rs$) $D_\ks$ becomes positive [cf.\ also Eq.\ \eqref{Ds}]. The vertical black line denotes the boundary between the paramagnetic electron gas ($p=0$, $\rs \lesssim 5.45$) and the ferromagnetic electron gas ($p=1$, $\rs \gtrsim 5.45$) in the exchange-only approximation. A negative magnon stiffness in a ferromagnet would imply that there is an instability with respect to the formation of a static spin wave. However, in the electron gas it is simply due to the fact that the electron gas at high densities ($\rs < 5.45$) is not ferromagnetic. This means that in order to stabilize a spin polarization an external magnetic field, $B_\mathrm{ext}$ is required, which--at the same time--causes the magnon dispersion to start at $\omega_m(q=0) = \Delta_\ks - \Delta_\x = 2 B_\mathrm{ext}$ at zero momentum transfer. Accordingly, the stability of the electron gas is not compromised if the curvature $D_\ks$ of the magnon dispersion is negative. Note that even if the previous discussion has been limited to the exchange-only approximation for the electron gas, nothing qualitatively changes if we consider correlation. In fact, the only substantial change is that including correlation moves the transition to the spin polarized electron gas to much smaller densities (larger $\rs$).

\begin{figure}[t]
  \begin{center}
    \includegraphics[width=0.95\linewidth,angle=0]{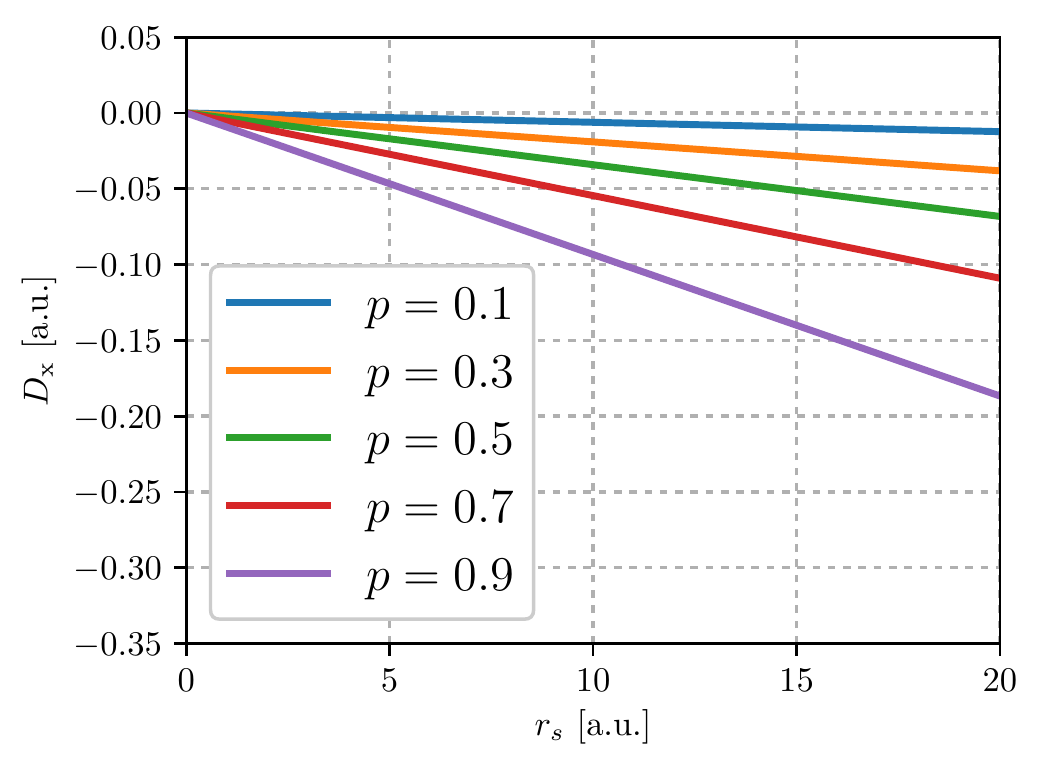}
    \caption{Plot of $D_{\x}$ vs. $r_s$ at polarization $p$. From Eq.\ \eqref{Ix2} we can see that $D_{\x}$ is a linear function of $\rs$ with a negative slope depending on the spin polarization $p$, since the only density dependence enters via the Fermi wave vector, which is inversely proportional to $\rs$ for the uniform electron gas.}
    \label{FIG:Dx}
  \end{center}
\end{figure}
In Fig.\ \ref{FIG:Dx} we plot the exchange correction to the magnon stiffness $D_\x$. From Eq.\ \eqref{Ix2} it is evident that $D_\x$ is proportional to $\rs$. At first sight it is surprising that the proportionality coefficient, which depends on the relative spin polarization $p$, is strictly negative, because the exchange energy tends to align spins. Hence, a transverse fluctuation--implying that spin are no longer perfectly aligned--is expected to have a positive contribution due to the exchange interaction. The reason why there is a gain in energy is the following: Equation \eqref{Ix2} has two contribution with opposite signs. The first term in the numerator, $\frac{8}{9} p^2$, corresponds to the first-order vertex and self-energy contributions, i.e., to the exchange energy and is, in fact, positive. However, as we have shown in Ref.\ \cite{EichVignale:13} there is an ``anomalous'' contribution to the first-order correction. This term arises because the perturbative expansion in SDFT is performed at constant density and spin magnetization, while standard perturbation theory is performed at constant chemical potential and magnetic field. The second term in the numerator of Eq.\ \eqref{Ix2}, $-\frac{2}{5}\lP \kp^5 - \km^5 \rP \lP \kp - \km \rP$, corresponds to this anomalous contribution, is negative and dominates the first term for all $p$ leading to an overall reduction of the magnon stiffness--at least at the level of the first-oder approximation. For a careful analysis and discussion of this anomalous term we refer the interested reader to Ref.\ \cite{EichVignale:13}. Comparing Figs.\ \ref{FIG:Ds} and \ref{FIG:Dx} we can see that the exchange correction to the magnon stiffness is relatively small compared to $D_\ks$. 

\section{Magnon dispersion in ferromagnets} \label{SEC:FMmagnons}
\begin{figure}[t]
  \begin{center}
    \includegraphics[width=0.95\linewidth,angle=0]{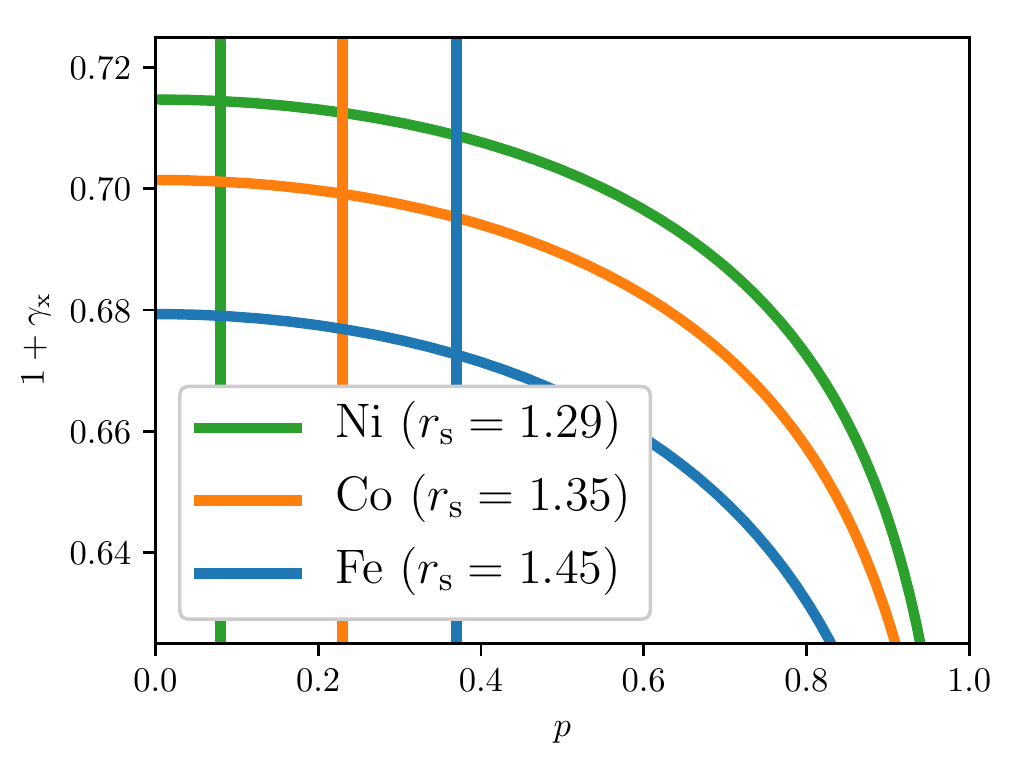}
    \caption{Plot shows the renormalization factor $1+\gamma_\x$ as function of the relative spin polarization $p$ for the \emph{ effective} $\rs$ corresponding to the ferromagnetic transition metals iron, cobalt and nickel. The vertical lines are at the relative spin polarizations of iron ($p=0.37$), cobalt ($p=0.23$) and nickel ($p=0.08$). \label{FIG:gamma}}
  \end{center}
\end{figure}
In the previous section, we have noted that the uniform gas does not polarize spontaneously at  densities relevant for ferromagnets \emph{ and} the corresponding transverse spin fluctuations start at finite frequency: $\omega_m(q=0) = \Delta_\ks - \Delta_\x = 2 B_\mathrm{ext}$.  Magnons in ferromagnets, instead,  have vanishing frequency for $q \to 0$. Thus, adopting the expressions derived in the previous section to estimate the dispersion of real magnons, we may additionally impose the condition $\Delta_\ks \equiv \Delta_\x$. This can be accommodated by replacing $\Delta_\x \to \Delta_\ks$ in the denominator of $D_\ks$ (see the first line in Eq.\ \eqref{Ds}),
\begin{align}
  D_\ks \to D_0 & = \frac{1}{2p} \lB 1 - \frac{ \kf^2\lP \kp^5 - \km^5 \rP}{5 \Delta_\ks} \rB \nn 
  & = \frac{1}{2p} \lB 1 - \frac{2 \lP \kp^5 - \km^5 \rP}{5 \lP \kp^2 - \km^2 \rP} \rB ~. \label{D0}
\end{align}
In the second line of Eq.\ \eqref{D0} we have used the explicit expression for Kohn-Sham spin splitting $\Delta_\ks = \frac{1}{2}\kf^2 \lP \kappa_+^2 - \kappa_-^2 \rP$.

It is reassuring to note that the aforementioned replacement is equivalent of adopting  a frozen-magnon approach, in which the magnon dispersion is obtained from energy differences from constrained ground-state calculations. This can be easily verified by
expanding  directly  the inverse of the \emph{ static} response function. For small $q$, we get $\omega  \approx D q^2$ with $D = D_0 + D_\x$. It should be clear that the overall procedure may be extended to include correlation contributions as well. Notice that $D_0$ depends only on the relative spin polarization $p$, so $D = D_0 + D_\x$ has the form of a perturbative expansion in $\rs$. The idea now is to use this expansion to define a renormalization factor for the magnon stiffness based on the results for the uniform electron gas, i.e.,
\begin{subequations}
  \begin{align}
    \frac{D}{D_0} & = 1 + \gamma_\x(\rs, p) = 1 + \frac{D_\x}{D_0}
    = 1 + \frac{d_\x(p)}{d_0(p)} \rs ~, \label{gamma} \\
    d_\x(p) & = \frac{\lB \frac{8}{9}p^2 - \frac{2}{5}\lP \kp^5 - \km^5 \rP \lP \kp - \km \rP \rB }
    {\lP 9 \pi / 4 \rP^{1/3} \pi \lB \kp^2 - \km^2\rB^2} ~, \label{dx} \\
    d_0(p) & = \frac{1}{2} - \frac{1}{5} \frac{\kp^5 - \km^5}{\kp^2 - \km^2} ~. \label{d0}
  \end{align}
\end{subequations}
Finally, we can use Eq.\ \eqref{gamma} to correct the magnon stiffness obtained from a calculation of $D$  employing the ALDA, i.e., $D \approx D_\mathrm{ALDA} (1 + \gamma_x)$. 

\begin{table}
  \begin{center}
    \begin{tabular}{l|c|c|c}
      $D$ [$\mathrm{meV\AA^2}$] & $D^\mathrm{exp}$&
      $D^\mathrm{ALDA}$ \cite{BuczekSandratskii:11} &
      $\tilde{D}$ \\
      \hline
      Fe (BCC) & $270-310$\cite{ShiranePickart:65,Stringfellow:68} & $250$ & $170$ \\
      Co (FCC) & $300-350$\cite{Hueller:86} & $490$ & $340$ \\
      Ni (FCC) & $400-550$\cite{Hueller:86} & $850$ & $610$ \\
    \end{tabular}
  \end{center}
  \caption{Comparison of the experimental results ($D^\mathrm{exp}$)for the magnon stiffness to the results obtained within ALDA ($D^\mathrm{ALDA}$) and the gradient-corrected $\tilde{D} = D^\mathrm{ALDA}(1 + \gamma_\x)$ (see Eq.\ \eqref{gamma} for the definition of $\gamma_\x$). Magnon stiffnesses are given in $\mathrm{meV\AA^2}$.}
  \label{TAB:D}
\end{table}
In Fig.\ \ref{FIG:gamma} we show the renormalization factor in the exchange approximation, $1 + \gamma_\x$, as function of $p$ for densities (specified in terms of $\rs$) corresponding to the ferromagnetic transition metals iron, cobalt and nickel. Since the spin magnetization is due to electrons in the 3d-bands we determine an \emph{ effective} density based on the number of 3d-electrons per atom and the unit cell volume of the transition metals. The relative spin polarization is determined by the spin magnetization per atom divided by the effective density. A comparison of the available experimental data for the magnon stiffness in iron, cobalt and nickel to linear-response TD-SDFT results obtained within the ALDA~\cite{BuczekSandratskii:11} and the ALDA results augmented by the renormalization $1 + \gamma_\x$ is presented in Table \ref{TAB:D}. We can see that the renormalization reduces the magnon stiffness obtained within the ALDA and that the reduction of $D$ due to the renormalization brings the computed magnon stiffness closer to the experimental results for cobalt and nickel. The ALDA result for iron is already in good agreement with the available experimental data and the renormalization moves the stiffness to smaller values. A possible explanation is that we are ignoring screening effects since we are using an exchange-only approximations. The d-electrons in iron are more delocalized compared to the d-electrons in cobalt and nickel, which could explain why screening plays an important role in iron, but is less important for cobalt and nickel.

\section{Conclusion and Outlook} \label{SEC:conclusions}

The ALDA has long been the workhorse of ab initio calculations of magnon dispersions in ferromagnetic materials.  The main merit of this approximation is to produce a reliable value for the magnon frequency at $q=0$. Specifically it guarantees that the magnon dispersion starts at the spin splitting energy due to the applied external magnetic field for zero momentum, which implies that in a ferromagnetic material it has zero energy at zero momentum.  The calculation of the frequency vs wave vector dispersion relation is less satisfactory, and usually results in an overestimation of the magnon stiffness.  In this paper we have argued that this overestimation most likely arises from the neglect of gradient corrections, which produce transverse components of the exchange-correlation fields. We have then devised a simple way to estimate these corrections by making use of analytic results from our previous work on the exchange-only energy of a non-uniform magnetized state in the homogeneous electron gas.   A key step in this estimate was the replacement of the complex ferromagnetic material by a homogeneous electron gas with reference density equal to the average density of the electrons that are responsible for magnetism (d-electrons in the case of the ``canonical'' metallic ferromagnets iron, cobalt and nickel) and an average spin polarization obtained from the equilibrium magnetization.  The gradient corrections to the magnon dispersion are found to be surprisingly large ($\sim 30\%$) and negative.  Thus they significantly improve the agreement between calculated and measured values of the magnon stiffness for cobalt and nickel. Admittedly, our empirical prescription falls shorts of the requirements for a true ab initio calculation and also does not work well for iron. However, it points unambiguously to the importance of gradient corrections, and suggests that more rigorous work able to capture more realistically the effects of magnetic inhomogeneities, based for example on the U(1)$\times$SU(2)-invariant functionals for non-collinear SDFT proposed in Ref.~\cite{PittalisEich:17} would go a long way in solving this long-standing problem in the ab initio theory of magnetism. 

\emph{Acknowledgements} -- F. G. E. has received funding from the European Union's Framework Programme for Research and Innovation Horizon 2020 (2014-2020) under the Marie Sk{\l}odowska-Curie Grant Agreement No. 701796.  GV was supported by the grant No. DE-FG02-05ER46203 funded by the U.S. Department of Energy, Office of Science.

\bibliography{magnonStiffness}

\end{document}